\documentclass[oneside,english]{amsart}
\pdfoutput=1
\usepackage[T1]{fontenc}
\usepackage[latin9]{inputenc}
\usepackage{geometry}
\usepackage{amssymb}
\usepackage{graphicx}

\makeatletter

\providecommand{\tabularnewline}{\\}

\makeatother

\usepackage{babel}
\begin{document}

\title{Corroborating pseudoscalar probing model with pulsar polarisation
datasets}

\author{Karam Chand \& Subhayan Mandal}

\address{Physics Department, Malaviya National Institute Of Technology, Jaipur,
Rajasthan-302017, India}

\email{2015rpy9054@mnit.ac.in, smandal.phy@mnit.ac.in}

\date{$\today$}
\begin{abstract}
Recently, we have used, pulsar polarisation datasets, on circular
polarisation degree \cite{parkesix18} \& linear polarisation position
angle \cite{ranki15}, to relate with well established theories, on
ellipticity parameter and linear polarisation position angles, accrued
by unpolarised photons, while undergoing photon-ALP oscillations,
inside a magnetised medium. This has given us parameter values such
as ALP mass and its coupling to photons \cite{mandal19}. To further
test this, we now switch to different wavebands, other than earlier
21 cm wavelength, and check for the validity of our model. Here we
use two data sets \cite{dmit16} on circular polarisation degree of
identical pulsars observed in two different wavebands. We show, correlation
between these two new sets of data and our model using the composite
product variable of ALP mass and its coupling to photons, exist. We
also check whether our model hypothesis that one physical effect,
namely ALP-photon mixing is sufficient to, estimate ALP parameters,
faithfully, or not. We conclude by describing other pertinent physical
effects that may be included into our model to explain the circular
polarisation degree of pulsars,independent of its operating wavelength
of observation. 
\end{abstract}

\keywords{ALP-$\gamma$ mixing, Pulsar, Polarisation }
\maketitle

\section{\label{sec:Intro}Introduction}

Generic pseudoscalars \cite{Maiani86,wilc78,pec77,wein78,kim79,sik83}
or Axion-like particles (ALPs) are well motivated by many throeries
beyond standard model of particle physics \cite{sengupta99,sen01}.
The proposed interconversion of photon to ALPs mediated by a background
magnetic field has been explored for more than two\cite{Das05,Das08,pand02,saral04}
decades to extract the mass \& couling strength to photons, of ALPs.
This field has been traversed by both phenomenologically \cite{payez11,payez13,payez14,payezcap12,payezcap15,payezprd12,vogel17,montani09,bass10,agarwal12,jain13,pelgrims15,pelgrims16,mandal16,tiwari12,tiwari16}
and observationally \cite{hut98,lamy00,hut01,hut05,hut10,jackson07,joshi07,jagan16}.
Here we shall turn our attention to the polarisation properties of
highly degenerate compact stars \cite{bell_68,pul_hbk_05,pul_mag_II_16},
commmonly referred to as neutron star. We have recently shown that
experimental observation of ellipticity parameters of such objects
\cite{mckinn02,ganga10,ardavan08,diagtool13} may also be related
to a theoretical ellipticity parameter which assumes a suitable $\gamma$-ALP
interconversion model, with the help of correlation and regression
we obtained the ALP mass \& its coupling to photons, in an earlier
work \cite{mandal19}. The composite product of these two pseudoscalar
parameters, may now be harnessed in reverse, to estimate waveband
frequencies in which the polarisation observations are made. In so
doing, we utilise another dataset describing polarisation details
of pulsars in two different wavebands, to calculate the observational
ellipticity parameter. Putting the composite product pseudoscalar
parameters and magnetic field, into the theoretical ellipticity parameter,
leaves only one unknown quantity, i.e. the frequency of observations,
aside. If our hypothesis is correct about this model of ALP-$\gamma$
mixing, then the slope of regression between theoretical and experimental
ellipticity will fit the value of the frequency of of observation,
in both the cases. Matches of frequencies shall increase the confidence
in our simple model. Otherwise possible improvements may be implemented
over it.

\section{\label{sec:DaMo}Data \& Model}

The following data shown in table no. (\ref{tab:Pul-Pol-Prop}) is
obtained from the reference no. \cite{dmit16}. It contains the spin
down luminosity ${\rm E}$, pulsar spin period ${\rm P}$ and spin
period time derivative ${\rm \dot{P}}$, for one hundred pulsars observed
in two frequencies, 333 MHz \& 618 MHz. Out of which 86 of them have
complete polarisation information, that are used here. However a fraction
of the data is only given here, for want of space. Following the basic
pulsar model \cite{pul_hbk_05} we may derive pulsar magnetic field
$\mathfrak{B}$. 
\begin{equation}
\mathfrak{B=10^{12}}{\rm \left(\dot{P}\right)_{-15}^{\frac{1}{2}}\left(P\right)_{sec}^{\frac{1}{2}}Gauss}\label{eq:mag-field}
\end{equation}
The ellipticity parameter can then be evaluated by the following formula
\cite{mandal19,ganguly12,cameron99}
\begin{equation}
\chi={\rm \frac{1}{96\omega}\left(g_{\phi\gamma\gamma}\mathfrak{B}m_{\phi}\right)^{2}z^{3}}\label{eq:ellpmtr}
\end{equation}
Given that the distance ${\rm z}$ is fixed to 10 kilometres \& the
magnetic field $\mathfrak{B}$ is calculated from the data, we can
calculate a theoretical estimate of the ellipticity parameter. In
so doing we use the values of pseudoscalar ALP parameters derived
in \cite{mandal19}, provided we use employ the ALP parameters as
derived therein, and the frequency used in \cite{dmit16}. However,
this also entails an unique opportunity to pit the the experimental
ellipticity parameter, as given by
\begin{equation}
\chi=\frac{1}{2}\arctan\frac{|\mathbb{V}|}{{\rm \mathbb{P}_{lin}}}\label{eq:ellpmtr-obs}
\end{equation}
with that of the theoretical one (\ref{eq:ellpmtr}), modulo the value
of the observational frequency, in what is known as a regression analysis.
The interceptless slope or the regression coefficient shall then reveal
to us the frequency used for the experiment. A close match shall boost
our confidence in our model. However, we shall look for other statistical
checks and balances such that how our results are self consistent
(closely correlated), and whether or not, the result thus obtained
is due to serendipitous stroke of luck, such as providential positioning
of outliers in the data. We shall repeat this method to chek our model
for both the wavebands of observation as given in our data source
reference \cite{dmit16}. The goodness of fit for both the cases shall
also inspected. Also, possible reasons for the differences shall be
attributed to effects that are already known in \cite{mandal09}.

\section{\label{sec:res}Result of Regression Analysis}

The result of the statistical analysis for the waveband 333 MHz is
given in table no. (\ref{tab:333})

\begin{table}[h]
\begin{tabular}{|c|c|c|c|c|c|}
\hline 
Coefficients & Mean & Std. Error & F-Statistics & t-value & Pr$\,\left(>|t|\right)$\tabularnewline
\hline 
\hline 
Slope & 2.050e-16 & 6.627e-17 & 9.5686 & 3.093 & 0.00268 $\ast \ast$\tabularnewline
\hline 
\end{tabular}

\caption{\label{tab:333}Result for 333MHz}
\end{table}

This is quite good fit with two stars at $4\sigma$ level. The above
result translates into a frequency of ${\rm 311.446\pm100.68\,MHz}$.
Except for the large WSSR error, the fit is quite convincing.

\begin{figure}[h]
\includegraphics[scale=0.5]{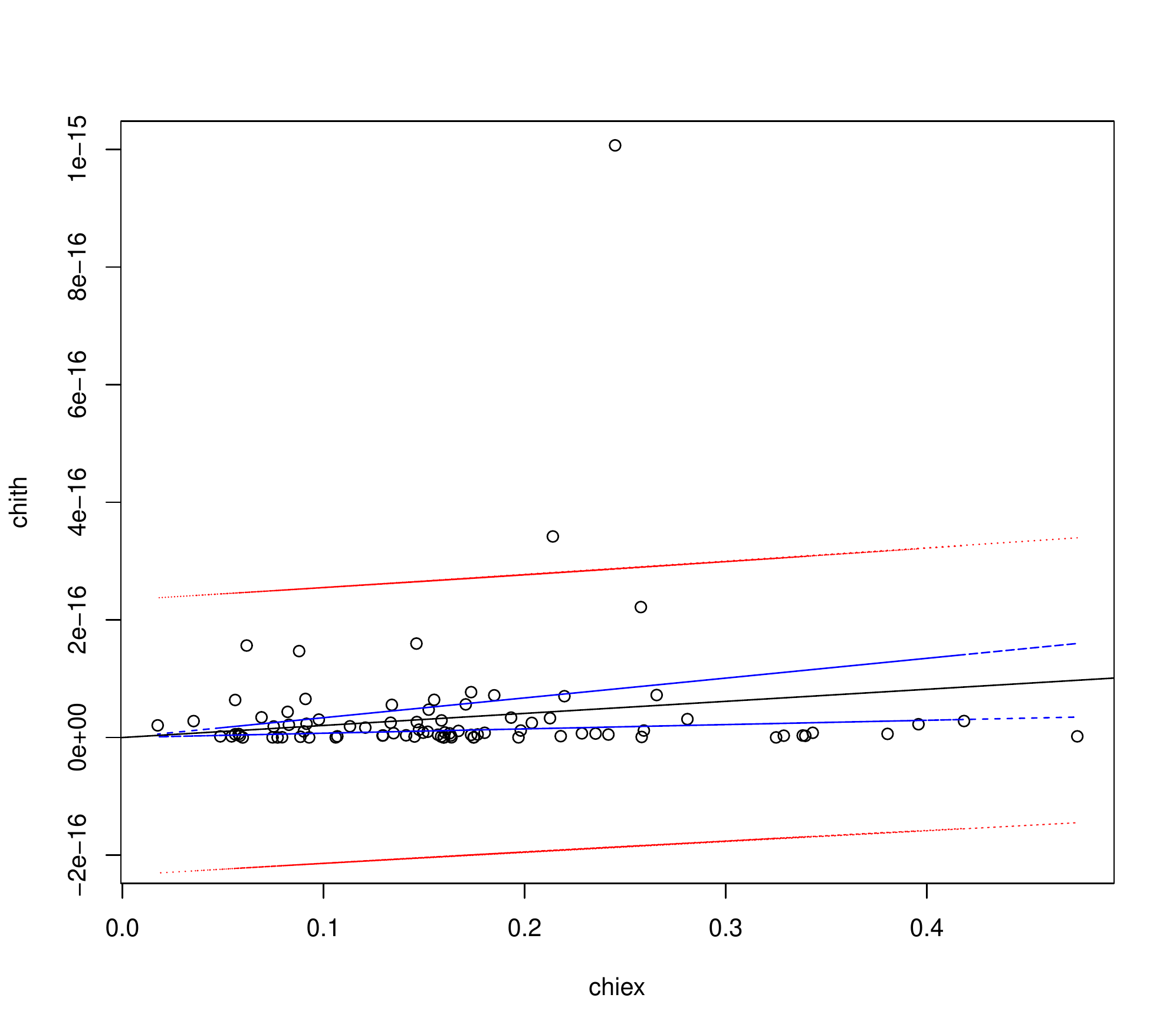}

\caption{\label{fig:333-sum}Summary fit graph for 333 MHz with confidence
interval in blue and prediction interval in red.}
\end{figure}

The summary graph containing the prediction limits and confidence
intervals (95\%) of the fit is given in fig. no. (\ref{fig:333-sum})

\begin{figure}
\begin{minipage}[t]{0.45\columnwidth}%
\includegraphics[scale=0.4]{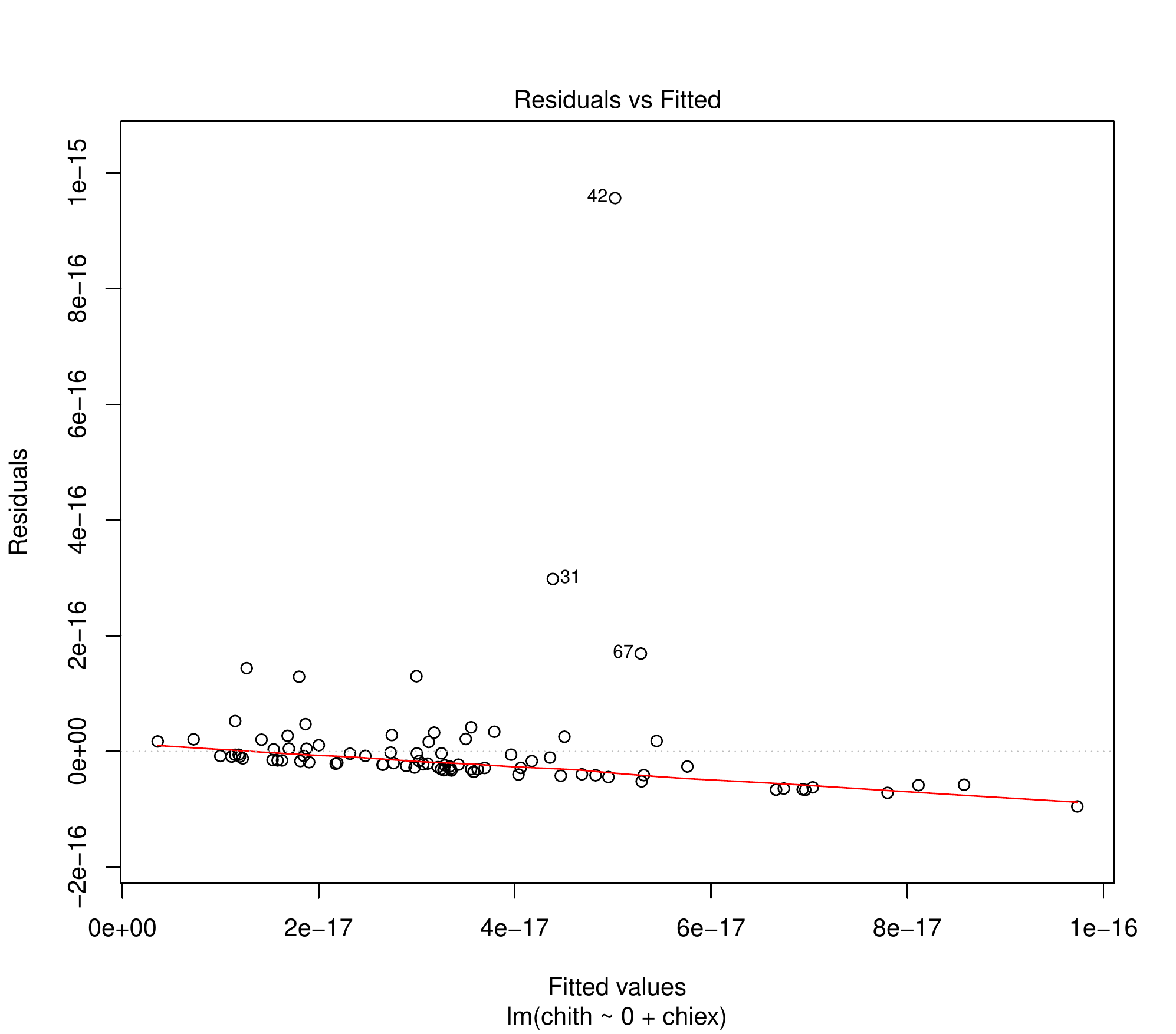}%
\end{minipage}%
\begin{minipage}[t]{0.45\columnwidth}%
\includegraphics[scale=0.4]{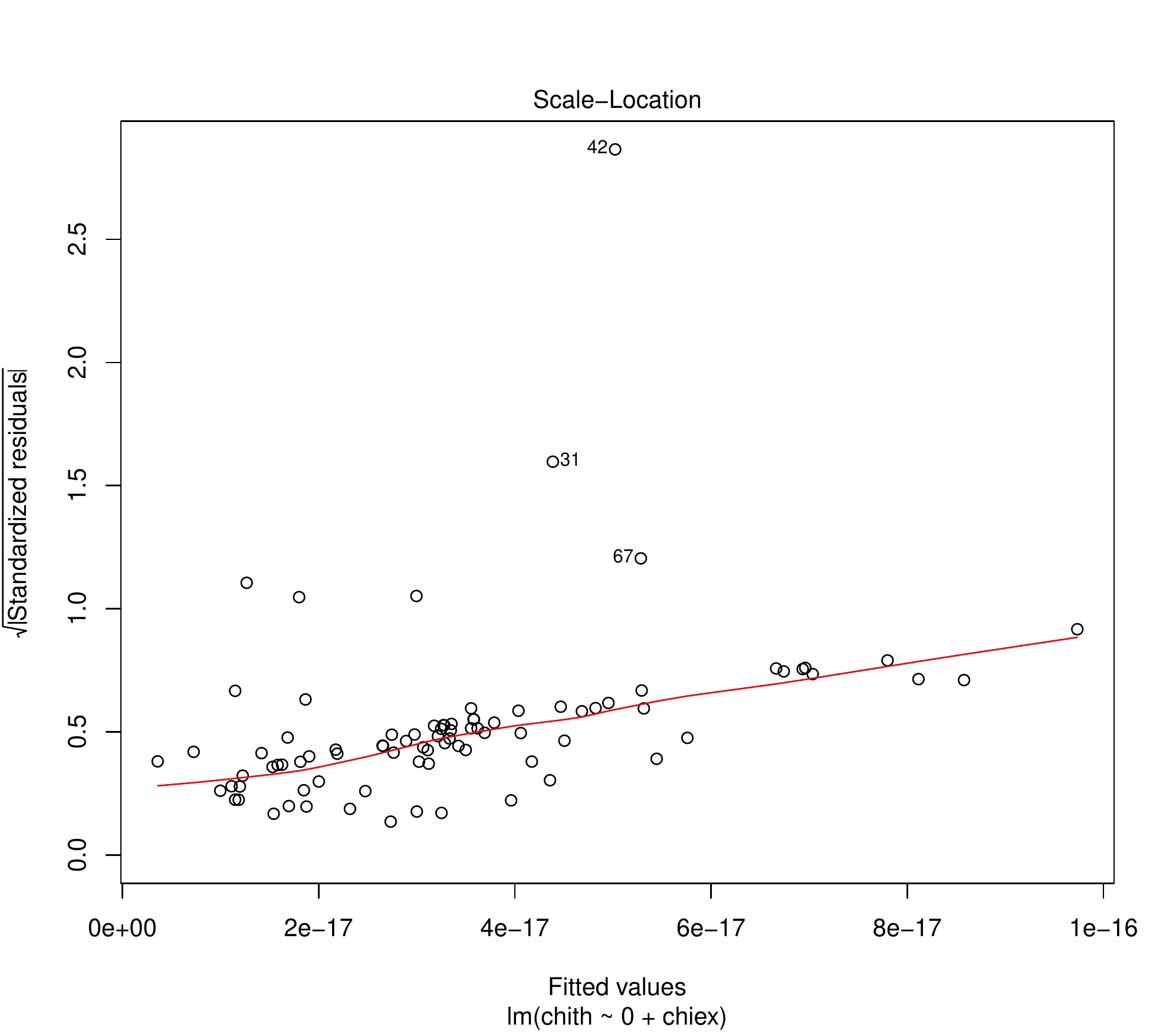}%
\end{minipage}

\begin{minipage}[t]{0.45\columnwidth}%
\includegraphics[scale=0.4]{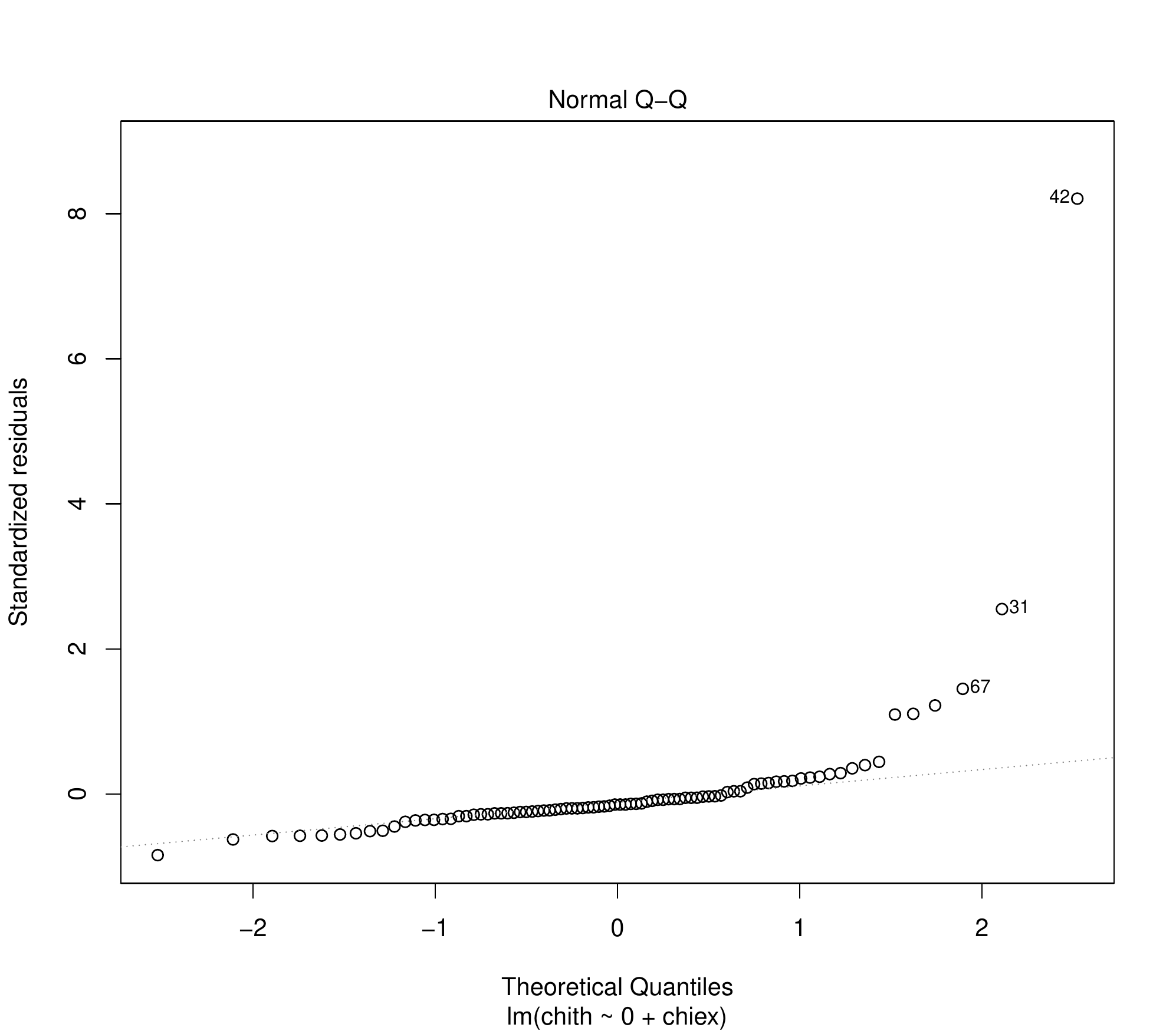}%
\end{minipage}%
\begin{minipage}[t]{0.45\columnwidth}%
\includegraphics[scale=0.45]{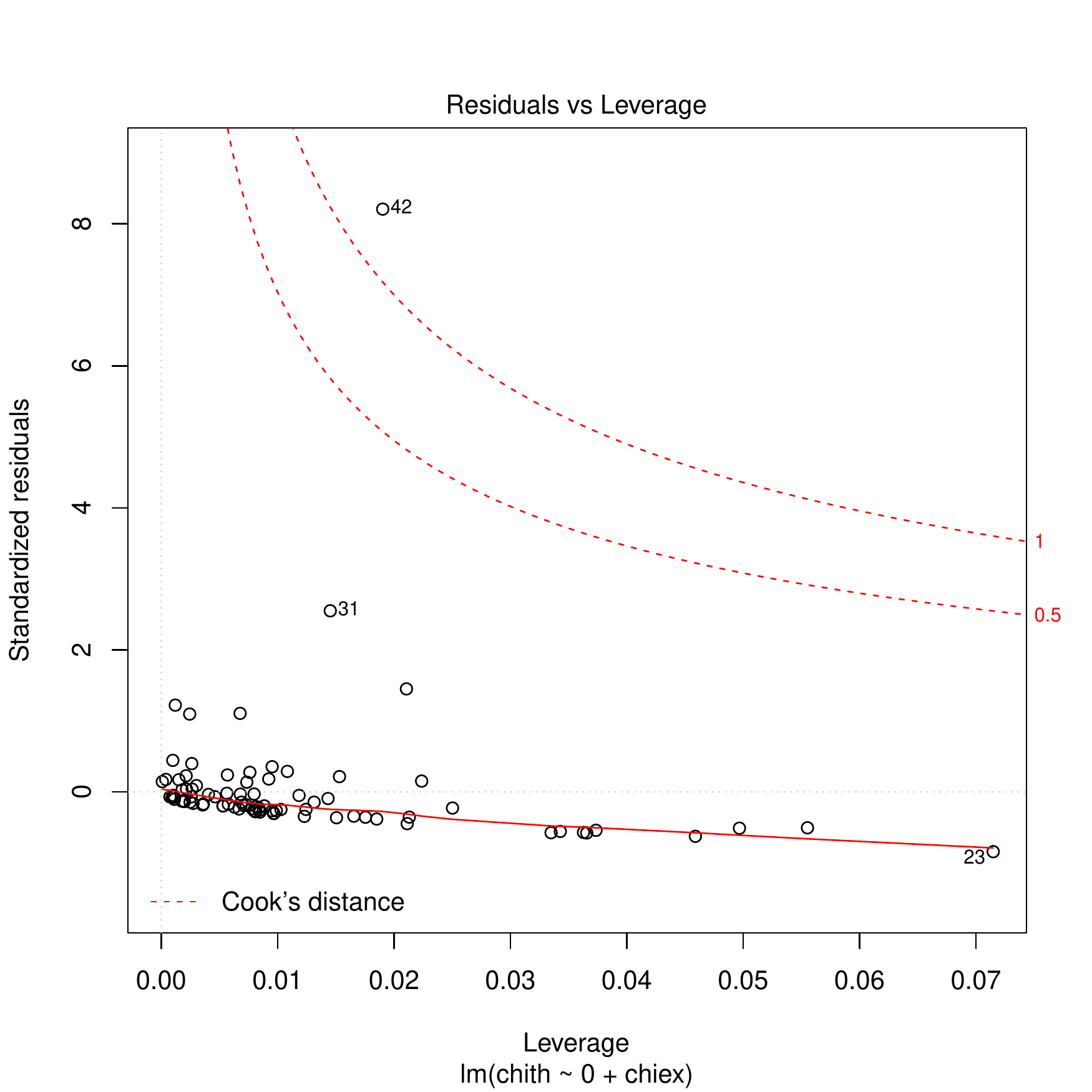}%
\end{minipage}

\caption{\label{fig:333-ch-fit}The fit characteristic curves for 333 MHz}
\end{figure}

Next, we look for other fitting characteristics for this waveband,
before going onto the other waveband. These include (standardized)
residues plot, Q-Q plot \& cook's distance plot, given in fig. no.
(\ref{fig:333-ch-fit}), so that we know the fit is not accidental
due to fortuitous placements of outliers. We also note the normal
nature of the distribution of the data points from Q-Q plot.

Now result of the statistical analysis for the waveband 618 MHz is
given in table no. (\ref{tab:618})

\begin{table}[h]
\begin{tabular}{|c|c|c|c|c|c|}
\hline 
Coefficients & Mean & Std. Error & F-Statistics & t-value & Pr$\,\left(>|t|\right)$\tabularnewline
\hline 
\hline 
Slope & 1.763e-16 & 6.692e-17 & 6.9422 & 2.635 & 0.01$\ast$\tabularnewline
\hline 
\end{tabular}

\caption{\label{tab:618}Result for 618MHz}
\end{table}

As compared to the earlier wavebands case, this is a poor fit, with
single star and at $3\sigma$ level. This result translates into $267.843\pm101.67{\rm MHz}$.
This, even with the large associated WSSR errorbars fails, quite markedly,
to reproduce the observational frequency. We shall discuss briefly,
the reasons behind this, in the next section.

\begin{figure}
\centering
\includegraphics[scale=0.5]{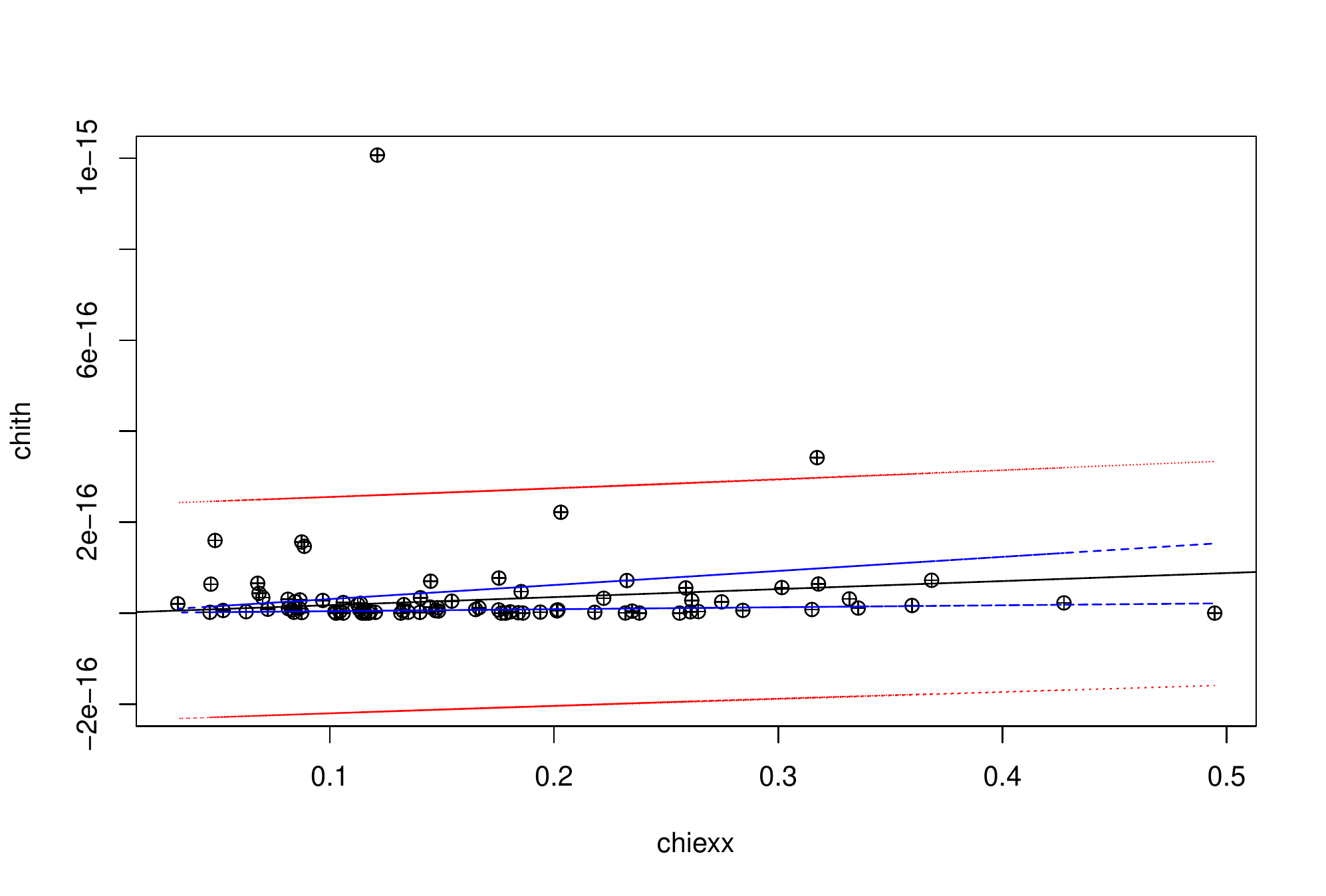}
\caption{\label{fig:618-sum}Summary fit graph for 618 MHz with confidence
interval in blue and prediction interval in red.}
\end{figure}

The summary graph containing the prediction limits and confidence
intervals (95\%) of the fit, for this waveband, too, is given in fig.
no. (\ref{fig:618-sum})

\begin{figure}
\begin{minipage}[t]{0.45\columnwidth}%
\includegraphics[scale=0.4]{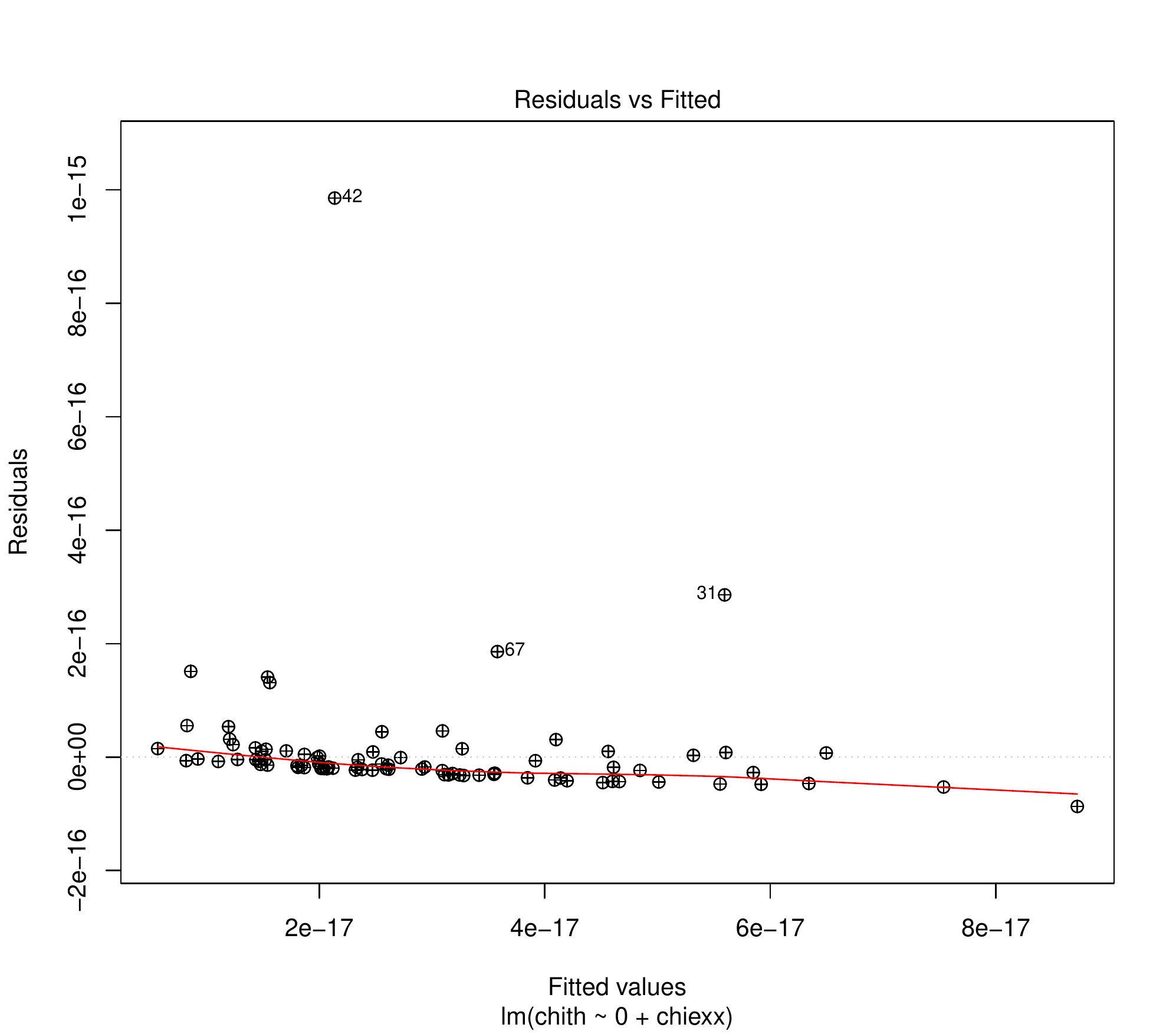}%
\end{minipage}%
\begin{minipage}[t]{0.45\columnwidth}%
\includegraphics[scale=0.4]{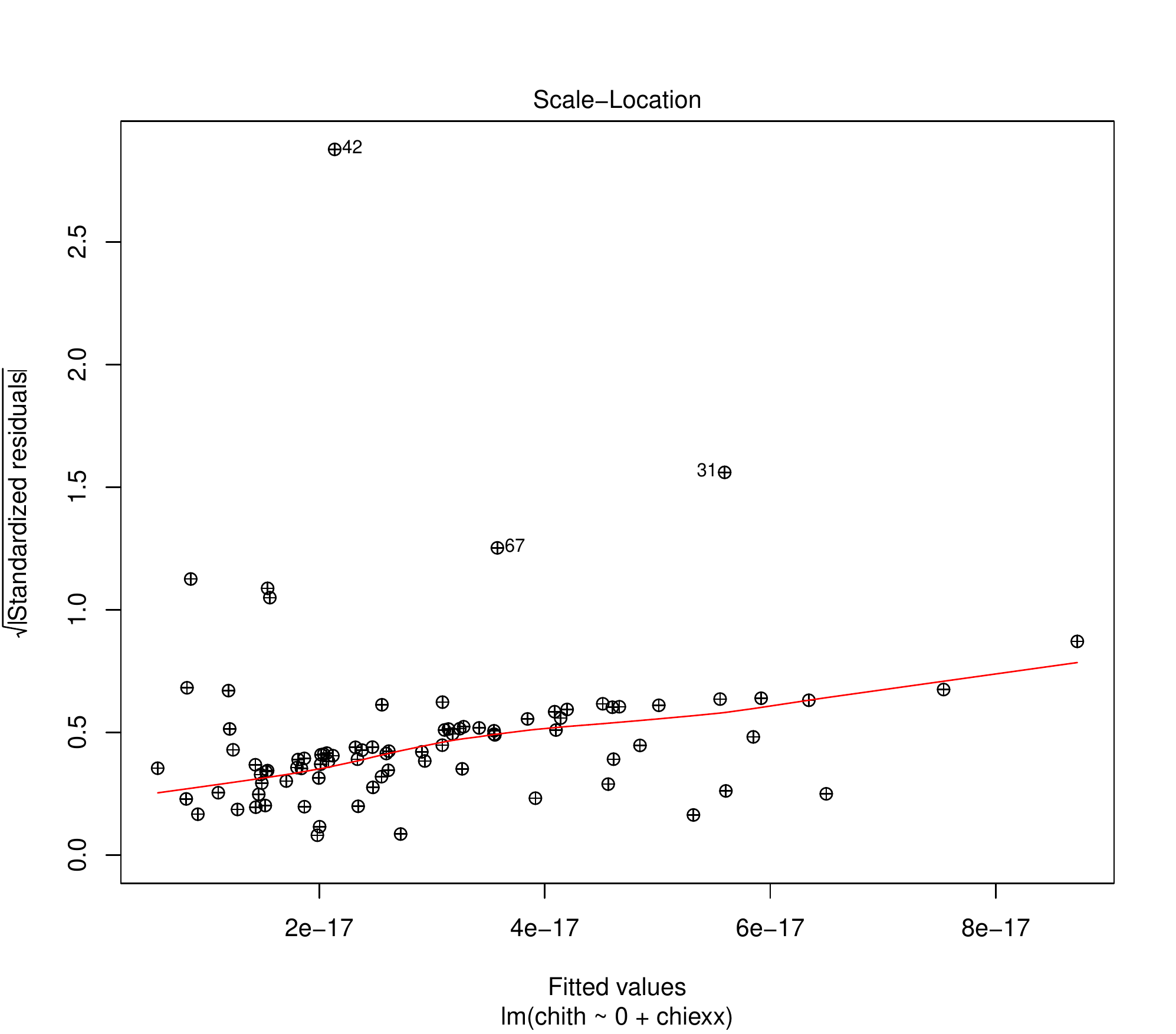}%
\end{minipage}

\begin{minipage}[t]{0.45\columnwidth}%
\includegraphics[scale=0.4]{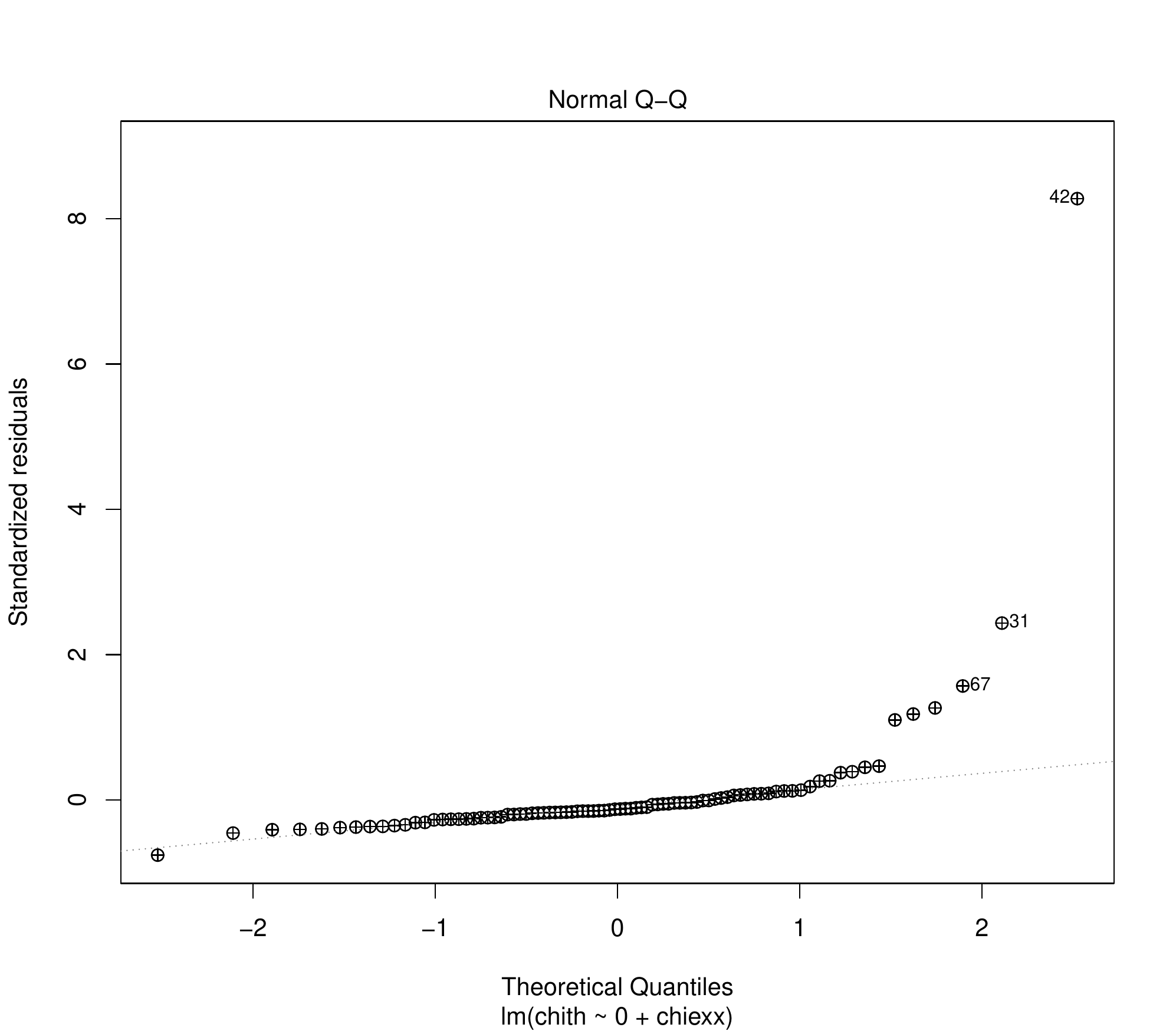}%
\end{minipage}%
\begin{minipage}[t]{0.45\columnwidth}%
\includegraphics[scale=0.45]{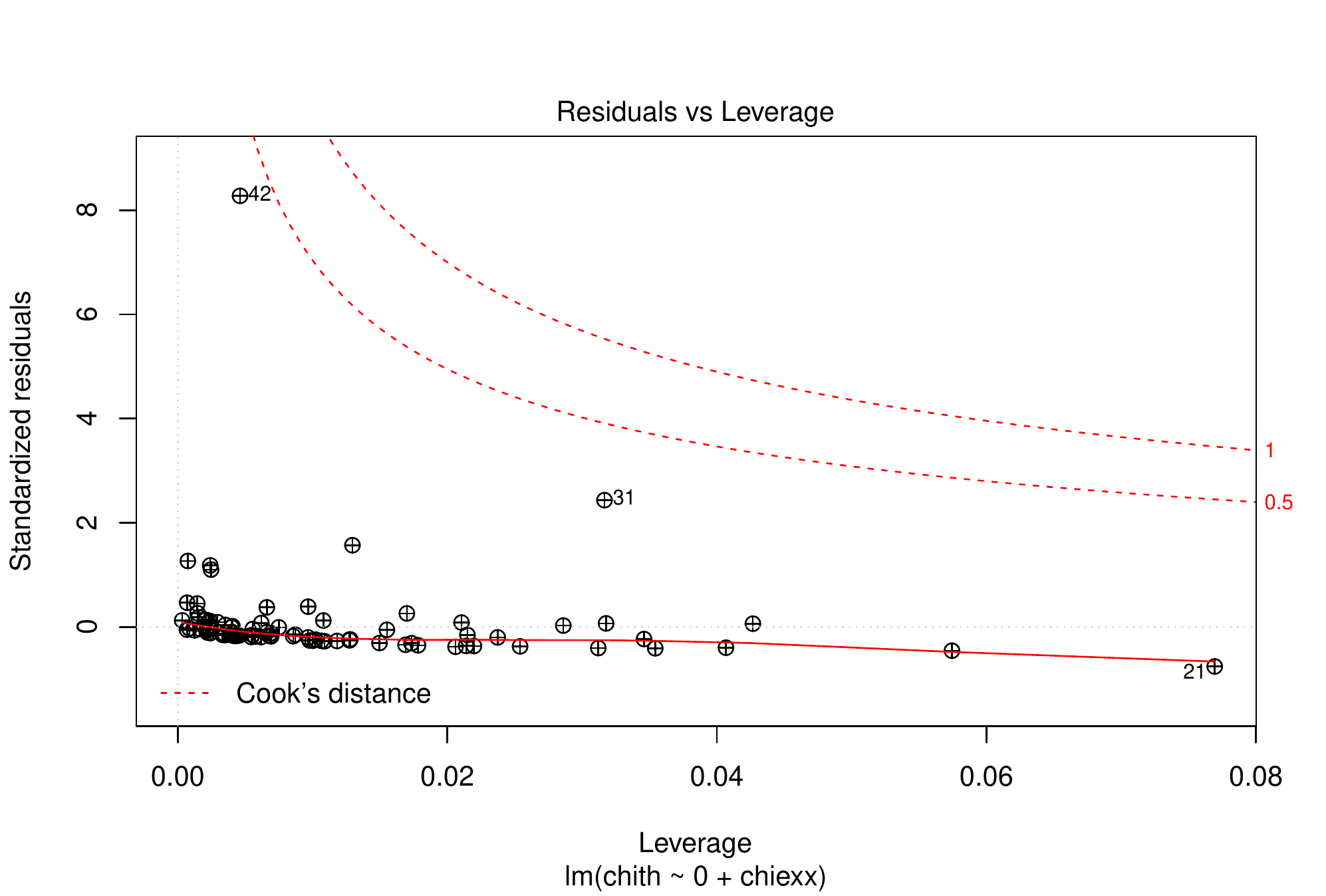}%
\end{minipage}

\caption{\label{fig:618-ch-fit}The fit characteristic curves for 618 MHz}
\end{figure}

Next, for the sake of completeness, we look for other fitting characteristics
for this waveband, too. These include, once again, (standardized)
residues plot, Q-Q plot \& cook's distance plot, given in fig. no.
(\ref{fig:618-ch-fit}), so that we know the fit is not accidental
due to fortuitous placements of outliers. We also note the normal
nature of the distribution of the data points from Q-Q plot.

\begin{figure}[b]
\includegraphics[scale=0.45]{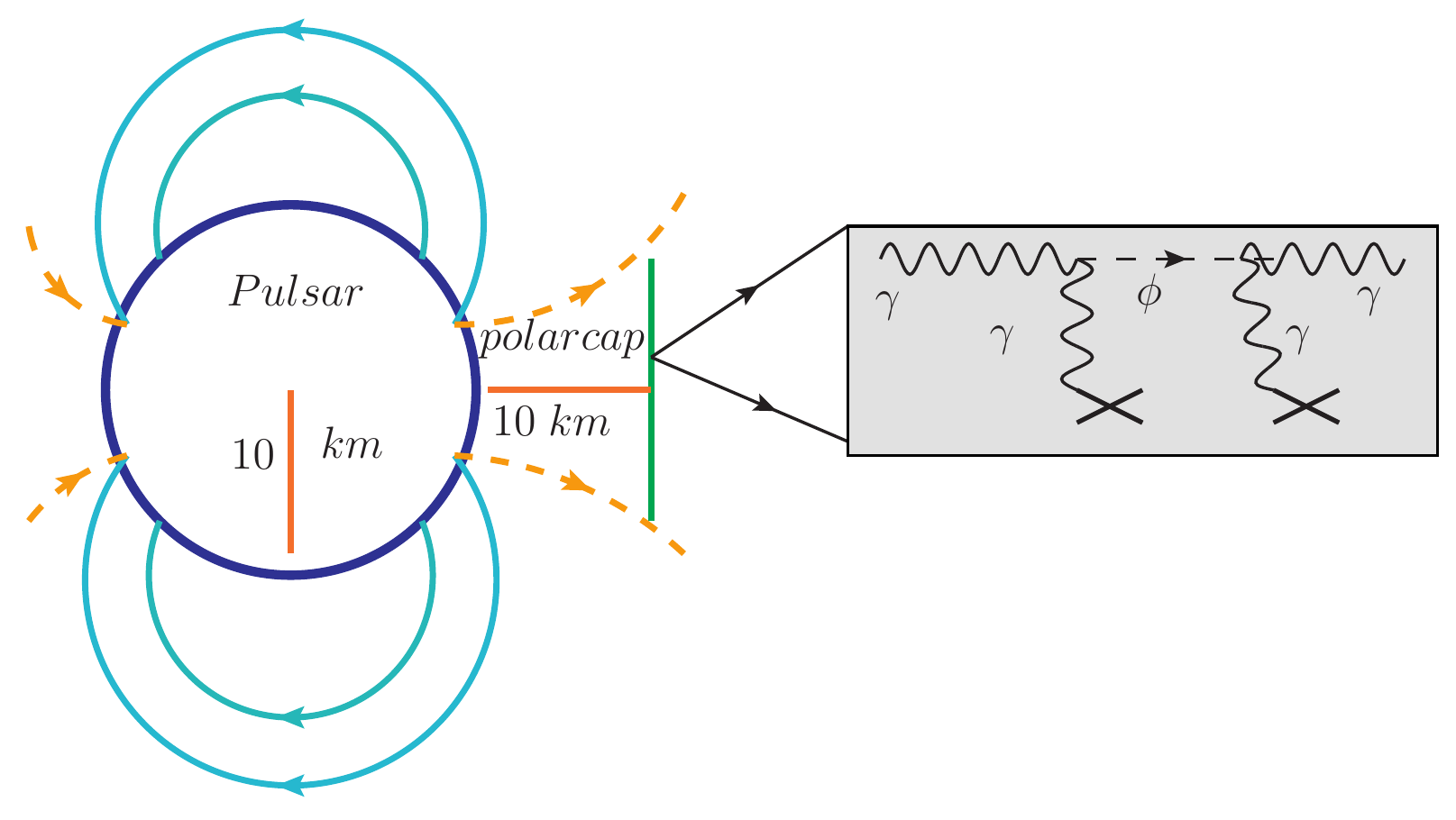}

\caption{\label{fig:interconversion}ALPs-$\gamma$ interconversion in the
polar cap region}

\end{figure}

\section{\label{sec:Dis}Discussion}

we have been successful in at least reproducing one observational
radio frequency, successfully, employing previously calculated values
of ALPs parameters. Also, the estimate veered off course for the other
observational frequency. The reason for this discrepancy requires
deeper investigation, whereupon we shall only underscore some tell
tale signature. We note that 618 MHz roughly translates to $4.067\times10^{-7}{\rm eV}.$
In case the limitting frequency as given in \cite{mandal09} evaluates
to this value, near the pulsar polar cap surroundings, as shown in
fig. no. (\ref{fig:interconversion}) then we may have to take into
account the Faraday effect, along side the mixing effect. Then we
have to redo all our calculation done previously in \cite{mandal19}
to accomodate for borderline frequencies. The limmiting frquency is
given by:
\begin{equation}
{\rm \omega_{L}=\left|\frac{\omega_{B}\omega_{P}\left(\omega_{p}^{2}-m_{\phi}^{2}\right)}{g_{\phi\gamma\gamma}^{2}\mathfrak{B}^{2}}\right|}^{\frac{1}{3}}\label{eq:limit}
\end{equation}
Assuming, the secondary plasma frequency as given by \cite{dmitra17}
we can actually compute this value quite accurately. The limitting
frequency indeed ${\rm \omega_{L}}$evaluates to $\sim{\rm 10^{-7\,}eV}$.
There are also other geometrical effects that may be important here.
As per the emission height to frequency mapping of pulsar radiation
\cite{gil03}, high frequency radio beams are generated parallel to
the magnetic axis \cite{maan13} whereas low frequency beams are generated
perpendicularly. Hence it is the intermediate frequency radio beam
that will more prone to exhibit Faraday type mixing between two photon
polarisations. It is also a known fact, that, Faraday effect is also
more pronounced at intermediate regions of frequency as given in by
the stokes parameter in ref. \cite{mandal09}, vide fig. no. 1 \&
2 therein.

\section{\label{sec:conclu}Conclusion \& Outlook}

The present analysis has proven to be insightful. On one hand it boosts
our confidence in the simple model that leads us to the ALPs parameters.
It also points to us the shortcoming of this simple model that may
creep up in borderline cases, on the other hand. We shall carefully
study and incorporate these effects and shall aspire to plug the theoretical
gaps in the ALPs parameter extraction model.

{\tiny{}}
\begin{table}
{\tiny{}}%
\begin{tabular}{lllllllllll}
 & \multicolumn{2}{l}{{\tiny{}333MHz}} & \multicolumn{2}{l}{{\tiny{}618 Mhz}} &  &  &  &  & \multicolumn{2}{l}{\;\;\;\;\;\;\;\;\;\;\;\;\;\;Unitless}\tabularnewline
{\tiny{}Jname} & {\tiny{}\%L} & {\tiny{}\%IVI} & {\tiny{}\%L} & {\tiny{}\%IVI} & {\tiny{}Period(s)} & {\tiny{}${\rm \dot{P}}$ -15(ss$^{-1}$)} & {\tiny{}$\mathfrak{B}$ (GeV$^{2}$)} & $\chi_{th}$(GeV) & $\chi_{Ex@333\,Hz}$ & $\chi_{Ex@6218\,Hz}$\tabularnewline
{\tiny{}J0034-0721} & {\tiny{}19.7} & {\tiny{}15.9} & {\tiny{}20.4} & {\tiny{}7.7} & {\tiny{}0.942} & {\tiny{}0.408} & {\tiny{}2.7405656827334E-08} & {\tiny{}2.84620291587993E-18} & {\tiny{}0.339529618702371} & {\tiny{}0.180458870791069}\tabularnewline
{\tiny{}J0134-2937} & {\tiny{}70.7} & {\tiny{}11} & {\tiny{}69.5} & {\tiny{}16.6} & {\tiny{}0.136} & {\tiny{}0.0784} & {\tiny{}4.56469923870566E-09} & {\tiny{}7.89604772133709E-20} & {\tiny{}0.077174732446713} & {\tiny{}0.117228154393456}\tabularnewline
{\tiny{}J0151-0635} & {\tiny{}33.9} & {\tiny{}11} & {\tiny{}38.9} & {\tiny{}11.9} & {\tiny{}1.464} & {\tiny{}0.4436} & {\tiny{}3.56246814269074E-08} & {\tiny{}4.80936133524069E-18} & {\tiny{}0.15688246056648} & {\tiny{}0.148436164504233}\tabularnewline
{\tiny{}J0152-1637} & {\tiny{}15.5} & {\tiny{}12.7} & {\tiny{}14.4} & {\tiny{}10.5} & {\tiny{}0.832} & {\tiny{}1.3} & {\tiny{}4.59746065840699E-08} & {\tiny{}8.009796307959E-18} & {\tiny{}0.34321587899093} & {\tiny{}0.315016954772958}\tabularnewline
{\tiny{}J0304+1932} & {\tiny{}39.5} & {\tiny{}12} & {\tiny{}37.6} & {\tiny{}11.2} & {\tiny{}1.387} & {\tiny{}1.3} & {\tiny{}5.93600952236433E-08} & {\tiny{}1.3352869566273E-17} & {\tiny{}0.147468529424404} & {\tiny{}0.144751835980694}\tabularnewline
{\tiny{}J0452-1759} & {\tiny{}24.3} & {\tiny{}4.5} & {\tiny{}15.8} & {\tiny{}3.4} & {\tiny{}0.548} & {\tiny{}5.75} & {\tiny{}7.84709452064903E-08} & {\tiny{}2.33347523727615E-17} & {\tiny{}0.091555408631242} & {\tiny{}0.10597881862131}\tabularnewline
{\tiny{}J0525+1115} & {\tiny{}23.1} & {\tiny{}9.6} & {\tiny{}25.8} & {\tiny{}14.5} & {\tiny{}0.354} & {\tiny{}0.0736} & {\tiny{}7.13551330788753E-09} & {\tiny{}1.92946040057403E-19} & {\tiny{}0.196934305368181} & {\tiny{}0.256010669791881}\tabularnewline
{\tiny{}J0543+2329} & {\tiny{}74.7} & {\tiny{}5.3} & {\tiny{}61.7} & {\tiny{}12.1} & {\tiny{}0.245} & {\tiny{}15.4} & {\tiny{}8.58673548957926E-08} & {\tiny{}2.79409776903932E-17} & {\tiny{}0.03541588634456} & {\tiny{}0.096826297710216}\tabularnewline
{\tiny{}J0614+2229} & {\tiny{}74.9} & {\tiny{}13.3} & {\tiny{}68.7} & {\tiny{}12.3} & {\tiny{}0.334} & {\tiny{}59.4} & {\tiny{}1.96902458246107E-07} & {\tiny{}1.46922295517181E-16} & {\tiny{}0.087869148342684} & {\tiny{}0.088581116851283}\tabularnewline
{\tiny{}J0629+2415} & {\tiny{}29.9} & {\tiny{}14.7} & {\tiny{}30.9} & {\tiny{}11.3} & {\tiny{}0.476} & {\tiny{}2} & {\tiny{}4.31323535550751E-08} & {\tiny{}7.05004260833669E-18} & {\tiny{}0.228468150184742} & {\tiny{}0.175294355633881}\tabularnewline
{\tiny{}J0630-2834} & {\tiny{}27.7} & {\tiny{}5.1} & {\tiny{}55.7} & {\tiny{}7.6} & {\tiny{}1.244} & {\tiny{}7.12} & {\tiny{}1.31563333598872E-07} & {\tiny{}6.5592648523076E-17} & {\tiny{}0.091038212022324} & {\tiny{}0.067803914675729}\tabularnewline
{\tiny{}J0659+1414} & {\tiny{}78.1} & {\tiny{}9.7} & {\tiny{}69.6} & {\tiny{}12.3} & {\tiny{}0.384} & {\tiny{}55} & {\tiny{}2.03156928978561E-07} & {\tiny{}1.56404306605116E-16} & {\tiny{}0.06178348632263} & {\tiny{}0.087459040244205}\tabularnewline
{\tiny{}J0729-1836} & {\tiny{}25.8} & {\tiny{}10} & {\tiny{}29.7} & {\tiny{}14.9} & {\tiny{}0.51} & {\tiny{}19} & {\tiny{}1.37608915060035E-07} & {\tiny{}7.1759362263427E-17} & {\tiny{}0.184884267396709} & {\tiny{}0.232496751604164}\tabularnewline
{\tiny{}J0738-4042} & {\tiny{}11.9} & {\tiny{}4.3} & {\tiny{}13.2} & {\tiny{}6.7} & {\tiny{}0.374} & {\tiny{}1.62} & {\tiny{}3.44094705772699E-08} & {\tiny{}4.48684854573428E-18} & {\tiny{}0.173372674007781} & {\tiny{}0.234844915770753}\tabularnewline
{\tiny{}J0742-2822} & {\tiny{}71.2} & {\tiny{}2.5} & {\tiny{}90} & {\tiny{}5.8} & {\tiny{}0.166} & {\tiny{}16.8} & {\tiny{}7.38232850940677E-08} & {\tiny{}2.06524777585392E-17} & {\tiny{}0.017548970232895} & {\tiny{}0.032177725823632}\tabularnewline
{\tiny{}J0758-1528} & {\tiny{}22.7} & {\tiny{}7} & {\tiny{}17.8} & {\tiny{}4.1} & {\tiny{}0.682} & {\tiny{}1.62} & {\tiny{}4.64659025065047E-08} & {\tiny{}8.18190028928016E-18} & {\tiny{}0.149558967455819} & {\tiny{}0.113194253747117}\tabularnewline
{\tiny{}J0837+0610} & {\tiny{}12.8} & {\tiny{}4.1} & {\tiny{}8.4} & {\tiny{}6.2} & {\tiny{}1.273} & {\tiny{}6.8} & {\tiny{}1.30062872220477E-07} & {\tiny{}6.41050302886615E-17} & {\tiny{}0.154993195623442} & {\tiny{}0.317919211761873}\tabularnewline
{\tiny{}J0905-5127} & {\tiny{}81.7} & {\tiny{}9.2} & {\tiny{}81.8} & {\tiny{}7.7} & {\tiny{}0.346} & {\tiny{}24.9} & {\tiny{}1.29754493042206E-07} & {\tiny{}6.38014045040587E-17} & {\tiny{}0.056067360888871} & {\tiny{}0.04692773430486}\tabularnewline
{\tiny{}J0908-1739} & {\tiny{}23.5} & {\tiny{}5.1} & {\tiny{}20.6} & {\tiny{}5.7} & {\tiny{}0.401} & {\tiny{}0.669} & {\tiny{}2.28965436044482E-08} & {\tiny{}1.98666794169735E-18} & {\tiny{}0.106853663344035} & {\tiny{}0.134972548802821}\tabularnewline
{\tiny{}J0922+0638} & {\tiny{}38.6} & {\tiny{}6.4} & {\tiny{}46.5} & {\tiny{}6.4} & {\tiny{}0.43} & {\tiny{}13.7} & {\tiny{}1.07294928238011E-07} & {\tiny{}4.3625841392554E-17} & {\tiny{}0.082154170738989} & {\tiny{}0.068387537344781}\tabularnewline
{\tiny{}J0944-1354} & {\tiny{}31.7} & {\tiny{}24.1} & {\tiny{}18.6} & {\tiny{}28.3} & {\tiny{}0.57} & {\tiny{}0.0453} & {\tiny{}7.1034808026488E-09} & {\tiny{}1.9121759473725E-19} & {\tiny{}0.325015199113481} & {\tiny{}0.49467281802259}\tabularnewline
{\tiny{}J0953+0755} & {\tiny{}33.1} & {\tiny{}5.3} & {\tiny{}17} & {\tiny{}8.5} & {\tiny{}0.253} & {\tiny{}0.23} & {\tiny{}1.06637231351906E-08} & {\tiny{}4.30926448927639E-19} & {\tiny{}0.079386544725543} & {\tiny{}0.231823804500403}\tabularnewline
{\tiny{}J1034-3224} & {\tiny{}6.8} & {\tiny{}9.5} & {\tiny{}20.8} & {\tiny{}9.7} & {\tiny{}1.15} & {\tiny{}0.23} & {\tiny{}2.27351341144054E-08} & {\tiny{}1.95875658603472E-18} & {\tiny{}0.474775908132418} & {\tiny{}0.218181966327288}\tabularnewline
{\tiny{}J1116-4122} & {\tiny{}5.1} & {\tiny{}1.4} & {\tiny{}6.5} & {\tiny{}3.7} & {\tiny{}0.943} & {\tiny{}7.95} & {\tiny{}1.21038656077139E-07} & {\tiny{}5.55179747146102E-17} & {\tiny{}0.13395521121167} & {\tiny{}0.258743870487054}\tabularnewline
{\tiny{}J1136+1551} & {\tiny{}31.8} & {\tiny{}14.4} & {\tiny{}25} & {\tiny{}11.9} & {\tiny{}1.187} & {\tiny{}3.73} & {\tiny{}9.30174560285111E-08} & {\tiny{}3.27879560386941E-17} & {\tiny{}0.212602509489809} & {\tiny{}0.222131965199621}\tabularnewline
{\tiny{}J1239+2453} & {\tiny{}46.6} & {\tiny{}14.6} & {\tiny{}46.8} & {\tiny{}7.7} & {\tiny{}1.382} & {\tiny{}0.96} & {\tiny{}5.09183498038968E-08} & {\tiny{}9.82503416946686E-18} & {\tiny{}0.151808913476855} & {\tiny{}0.081534478854709}\tabularnewline
{\tiny{}J1257-1027} & {\tiny{}25} & {\tiny{}8.2} & {\tiny{}25.7} & {\tiny{}7.4} & {\tiny{}0.617} & {\tiny{}0.363} & {\tiny{}2.09209102706742E-08} & {\tiny{}1.65861879520145E-18} & {\tiny{}0.158471449564976} & {\tiny{}0.140177049423554}\tabularnewline
{\tiny{}J1328-4357} & {\tiny{}34.7} & {\tiny{}19.8} & {\tiny{}29.2} & {\tiny{}10.1} & {\tiny{}0.532} & {\tiny{}3.01} & {\tiny{}5.59401654101237E-08} & {\tiny{}1.1858586375611E-17} & {\tiny{}0.259262594573903} & {\tiny{}0.166504517338386}\tabularnewline
{\tiny{}J1328-4921} & {\tiny{}23.2} & {\tiny{}11.8} & {\tiny{}21.6} & {\tiny{}9.2} & {\tiny{}1.478} & {\tiny{}0.61} & {\tiny{}4.19746242214031E-08} & {\tiny{}6.67665694834474E-18} & {\tiny{}0.235260176510989} & {\tiny{}0.201327326775936}\tabularnewline
{\tiny{}J1507-4352} & {\tiny{}50.2} & {\tiny{}17} & {\tiny{}34.9} & {\tiny{}8.4} & {\tiny{}0.286} & {\tiny{}1.6} & {\tiny{}2.990389025127E-08} & {\tiny{}3.38875997644419E-18} & {\tiny{}0.163261893917659} & {\tiny{}0.118097552618243}\tabularnewline
{\tiny{}J1527-3931} & {\tiny{}32} & {\tiny{}14.6} & {\tiny{}27.3} & {\tiny{}20.1} & {\tiny{}2.417} & {\tiny{}19.1} & {\tiny{}3.00358336996861E-07} & {\tiny{}3.41873006303656E-16} & {\tiny{}0.21401962844742} & {\tiny{}0.317325891138513}\tabularnewline
{\tiny{}J1555-3134} & {\tiny{}16.9} & {\tiny{}5.6} & {\tiny{}15.7} & {\tiny{}3.7} & {\tiny{}0.518} & {\tiny{}0.0622} & {\tiny{}7.93496023123998E-09} & {\tiny{}2.38602471453324E-19} & {\tiny{}0.159987178665129} & {\tiny{}0.115722827378211}\tabularnewline
{\tiny{}J1559-4438} & {\tiny{}47.1} & {\tiny{}4.6} & {\tiny{}53.1} & {\tiny{}9} & {\tiny{}0.257} & {\tiny{}1.02} & {\tiny{}2.26334797904343E-08} & {\tiny{}1.94127958965271E-18} & {\tiny{}0.048677894329502} & {\tiny{}0.083947961625188}\tabularnewline
{\tiny{}J1603-2531} & {\tiny{}52.9} & {\tiny{}14} & {\tiny{}38.1} & {\tiny{}4.8} & {\tiny{}0.283} & {\tiny{}1.59} & {\tiny{}2.96535342573529E-08} & {\tiny{}3.33225595848031E-18} & {\tiny{}0.129359466557577} & {\tiny{}0.062661993245773}\tabularnewline
{\tiny{}J1700-3312} & {\tiny{}43.9} & {\tiny{}13.8} & {\tiny{}40.4} & {\tiny{}15.7} & {\tiny{}1.358} & {\tiny{}4.71} & {\tiny{}1.11800828216467E-07} & {\tiny{}4.73669553892762E-17} & {\tiny{}0.152285057190668} & {\tiny{}0.185326184367988}\tabularnewline
{\tiny{}J1703-3241} & {\tiny{}43.9} & {\tiny{}5.1} & {\tiny{}52.3} & {\tiny{}5.5} & {\tiny{}1.211} & {\tiny{}0.66} & {\tiny{}3.95210724907106E-08} & {\tiny{}5.91892547808738E-18} & {\tiny{}0.057827340409833} & {\tiny{}0.052388703265948}\tabularnewline
{\tiny{}J1709-1640} & {\tiny{}27.8} & {\tiny{}5.5} & {\tiny{}12.8} & {\tiny{}2.1} & {\tiny{}0.653} & {\tiny{}6.31} & {\tiny{}8.97337779138269E-08} & {\tiny{}3.0513872967089E-17} & {\tiny{}0.097659719132651} & {\tiny{}0.081306914298975}\tabularnewline
{\tiny{}J1709-4429} & {\tiny{}63.8} & {\tiny{}30} & {\tiny{}82.8} & {\tiny{}24.7} & {\tiny{}0.102} & {\tiny{}93} & {\tiny{}1.36152695080193E-07} & {\tiny{}7.02486388473549E-17} & {\tiny{}0.219770302434653} & {\tiny{}0.144952430674232}\tabularnewline
{\tiny{}J1720-2933} & {\tiny{}20.8} & {\tiny{}16.7} & {\tiny{}19} & {\tiny{}11.1} & {\tiny{}0.62} & {\tiny{}0.746} & {\tiny{}3.00642199970663E-08} & {\tiny{}3.42519507059652E-18} & {\tiny{}0.338248690349166} & {\tiny{}0.264364342256138}\tabularnewline
{\tiny{}J1722-3207} & {\tiny{}7.3} & {\tiny{}3.4} & {\tiny{}19.1} & {\tiny{}7.8} & {\tiny{}0.477} & {\tiny{}0.646} & {\tiny{}2.45391847751143E-08} & {\tiny{}2.28194771997698E-18} & {\tiny{}0.217938486203693} & {\tiny{}0.193853486367456}\tabularnewline
{\tiny{}J1722-3712} & {\tiny{}44.3} & {\tiny{}10.2} & {\tiny{}43.7} & {\tiny{}10} & {\tiny{}0.236} & {\tiny{}10.9} & {\tiny{}7.0901256659103E-08} & {\tiny{}1.90499260563921E-17} & {\tiny{}0.113152103354616} & {\tiny{}0.112479846397991}\tabularnewline
{\tiny{}J1731-4744} & {\tiny{}15} & {\tiny{}8} & {\tiny{}18.2} & {\tiny{}4.5} & {\tiny{}0.829} & {\tiny{}164} & {\tiny{}5.15446828227316E-07} & {\tiny{}1.00682310174267E-15} & {\tiny{}0.244978663126864} & {\tiny{}0.121195673389309}\tabularnewline
{\tiny{}J1733-2228} & {\tiny{}22.5} & {\tiny{}8.2} & {\tiny{}25.4} & {\tiny{}13.1} & {\tiny{}0.871} & {\tiny{}0.0427} & {\tiny{}8.52526126559884E-09} & {\tiny{}2.75423392517306E-19} & {\tiny{}0.174742262345038} & {\tiny{}0.238083258387332}\tabularnewline
{\tiny{}J1735-0724} & {\tiny{}20.7} & {\tiny{}6} & {\tiny{}24.1} & {\tiny{}5.6} & {\tiny{}0.419} & {\tiny{}1.21} & {\tiny{}3.14763509508964E-08} & {\tiny{}3.75451796428636E-18} & {\tiny{}0.141061866071358} & {\tiny{}0.114156769706462}\tabularnewline
{\tiny{}J1740+1311} & {\tiny{}26.8} & {\tiny{}8.9} & {\tiny{}40.5} & {\tiny{}5.9} & {\tiny{}0.803} & {\tiny{}1.45} & {\tiny{}4.7700887129696E-08} & {\tiny{}8.62260200737061E-18} & {\tiny{}0.16031536701803} & {\tiny{}0.072330693788357}\tabularnewline
{\tiny{}J1741-0840} & {\tiny{}39.5} & {\tiny{}5.5} & {\tiny{}40.4} & {\tiny{}5.7} & {\tiny{}2.043} & {\tiny{}2.27} & {\tiny{}9.51988701727914E-08} & {\tiny{}3.43438530471096E-17} & {\tiny{}0.069175485088563} & {\tiny{}0.070081977043168}\tabularnewline
{\tiny{}J1741-3927} & {\tiny{}12.3} & {\tiny{}3.4} & {\tiny{}18.5} & {\tiny{}3.1} & {\tiny{}0.512} & {\tiny{}1.93} & {\tiny{}4.39438716211487E-08} & {\tiny{}7.31782573934242E-18} & {\tiny{}0.134844230870457} & {\tiny{}0.083012550689768}\tabularnewline
{\tiny{}J1745-3040} & {\tiny{}27.1} & {\tiny{}8.9} & {\tiny{}41.7} & {\tiny{}7.3} & {\tiny{}0.367} & {\tiny{}10.7} & {\tiny{}8.76011088002886E-08} & {\tiny{}2.90806852086947E-17} & {\tiny{}0.15865799683273} & {\tiny{}0.08665191821248}\tabularnewline
{\tiny{}J1748-1300} & {\tiny{}21.5} & {\tiny{}5.7} & {\tiny{}25} & {\tiny{}5.3} & {\tiny{}0.394} & {\tiny{}1.21} & {\tiny{}3.05228778260504E-08} & {\tiny{}3.53050137930508E-18} & {\tiny{}0.129577188270801} & {\tiny{}0.104453473491318}\tabularnewline
{\tiny{}J1750-3503} & {\tiny{}58.5} & {\tiny{}21.3} & {\tiny{}46.7} & {\tiny{}17.2} & {\tiny{}0.684} & {\tiny{}0.0381} & {\tiny{}7.13633486999034E-09} & {\tiny{}1.92990473099052E-19} & {\tiny{}0.174591347807496} & {\tiny{}0.176445575971771}\tabularnewline
{\tiny{}J1751-4657} & {\tiny{}33.2} & {\tiny{}11.2} & {\tiny{}26.7} & {\tiny{}11.4} & {\tiny{}0.742} & {\tiny{}1.29} & {\tiny{}4.32495397372966E-08} & {\tiny{}7.08840313429382E-18} & {\tiny{}0.162680323508321} & {\tiny{}0.201767467507319}\tabularnewline
{\tiny{}J1752-2806} & {\tiny{}8.6} & {\tiny{}3.5} & {\tiny{}10.4} & {\tiny{}3} & {\tiny{}0.562} & {\tiny{}8.13} & {\tiny{}9.44926683002232E-08} & {\tiny{}3.38362055462677E-17} & {\tiny{}0.19325315255696} & {\tiny{}0.140418862310602}\tabularnewline
{\tiny{}J1801-0357} & {\tiny{}16} & {\tiny{}16.2} & {\tiny{}19.4} & {\tiny{}22.3} & {\tiny{}0.921} & {\tiny{}3.31} & {\tiny{}7.71842154822863E-08} & {\tiny{}2.2575761966324E-17} & {\tiny{}0.395804631825157} & {\tiny{}0.427415368453832}\tabularnewline
{\tiny{}J1801-2920} & {\tiny{}34.5} & {\tiny{}10.4} & {\tiny{}37} & {\tiny{}11.8} & {\tiny{}1.081} & {\tiny{}3.29} & {\tiny{}8.33672504547439E-08} & {\tiny{}2.63376113824825E-17} & {\tiny{}0.146392936985865} & {\tiny{}0.154360987829776}\tabularnewline
{\tiny{}J1807-0847} & {\tiny{}34.1} & {\tiny{}4.1} & {\tiny{}18.7} & {\tiny{}3.9} & {\tiny{}0.163} & {\tiny{}0.0288} & {\tiny{}3.02882881507688E-09} & {\tiny{}3.47644117863191E-20} & {\tiny{}0.059830096679633} & {\tiny{}0.102804463758326}\tabularnewline
{\tiny{}J1808-0813} & {\tiny{}30} & {\tiny{}11.3} & {\tiny{}32.4} & {\tiny{}11.1} & {\tiny{}0.876} & {\tiny{}1.24} & {\tiny{}4.60731152190082E-08} & {\tiny{}8.0441578601677E-18} & {\tiny{}0.180115528117463} & {\tiny{}0.165030306204717}\tabularnewline
\end{tabular}{\tiny\par}

{\tiny{}\caption{\label{tab:Pul-Pol-Prop}Pulsar Polarisation Properties }
}{\tiny\par}
\end{table}
{\tiny\par}

\bibliographystyle{unsrt}
\addcontentsline{toc}{section}{\refname}\bibliography{emeq,test}

\end{document}